\documentclass[12pt,a4paper]{article}
\usepackage[english]{babel}
\usepackage[dvips]{graphicx}
\usepackage{latexsym,amsfonts,amssymb}
\usepackage{amsmath}
\usepackage{eufrak}
\usepackage{epsf}

\usepackage{xypic}
\usepackage{amscd}
\usepackage[all]{xy}
\usepackage{textcomp} 
\usepackage{makeidx} 
\usepackage{mathrsfs} 

\def\eq{\begin{equation}}
\def\en{\end{equation}}

\def\>>{\rangle}
\def\<<{\langle}

\def\sT{\mathscr T}
\def\sN{\mathscr N}

\def\somma{{\rm Sum}}
\def\rmMNC{\rm MNC}
\def\rmVASM{\rm VASM}
\def\rmCBC{\rm CBC}

\newtheorem{proposition}{\emph {Proposition}}

\def\sk{\vskip .4cm}

\begin{document}

\begin{titlepage}

\vskip .8truecm
\begin{center}
\Large\bf  The Rotor Model with spectral parameters and enumerations of
Alternating Sign Matrices.

\end{center}

\vskip 4truecm
\begin{center}
Luigi Cantini 
\footnote{Laboratoire de Physique Th\'eorique et Mod\`eles Statistiques,
  (UMR 8626 du CNRS) Universit\'e Paris-Sud, B\^atiment 100, 91405 Orsay
  Cedex, France. \\ e-mail: luigi.cantini@lptms.u-psud.fr}
\vskip .8truecm

\begin{abstract}

In this paper we study the Rotor Model of Martins and Nienhuis. After
introducing spectral parameters, 
a combined use of integrability,
polynomiality of the ground state wave function and a mapping into the
fully-packed $O(1)$-model allows us to determine the sum rule and a
family of maximally nested components for different boundary conditions. We
see in this way 
the appearance of 3-enumerations of Alternating Sign Matrices.

\end{abstract}
\bigskip

\vskip 1truecm
\end{center}

\end{titlepage}

\section{\bf Introduction}
In recent years it has become clear that some exactly integrable
models present a deep and somehow unexpected combinatorial structure,
related to 
various enumerations of alternating sign matrices (ASM).  
Such matrices appeared for the first time in the work of Mills,
Robbins and Rumsey \cite{MRR}, and from that moment on they have played a
fundamental 
role in modern developments of combinatorics \cite{bressoud}. 
An ASM is a matrix with only $0, 1, -1$ entries, such that on each row
and on each column $1$s and $-1$s appear alternately, possibly
separated by zeroes and the sum on each row and each column is equal to
$1$. 

In a series of remarkable papers \cite{raz-strog1} Razumov and Stroganov
noticed the appearance 
of various enumerations of ASMs in the components of the ground state of the
XXZ spin chain at $\Delta = -1/2$. After  \cite{raz-strog1} many other
models with different boundary conditions have been object of investigation,
leading to a plethora of conjectures (for a review see
\cite{degier}).

A first step towards the proof of these conjectures was made by Di Francesco and
Zinn-Justin in \cite{pdf-pzj-1}. Their idea, first applied to the fully
packed $O(1)$ 
loop-model \cite{raz-strogO(1)_1, batch1}, was to introduce spectral
parameters, so that the 
components of the ground state become polynomials in these. 
Then, making use of the integrability of the model, they were able to
prove a set of equations satisfied by these polynomials, which were
generalised and reinterpreted as qKZ equations \cite{pasquier,
  pdf-pzj-2}. 
In that way they were not only able to prove a conjecture related to the sum
rule, but they also discovered the deep role of algebraic geometry in the
game \cite{knutson}. 

\sk

In the present article we apply the idea of Di Francesco and Zinn-Justin to
the study of a model introduced by Martins and Nienhuis \cite{martins-1998}
and called rotor model. In \cite{rotor} Batchelor, de Gier and Nienhuis have
computed numerically the components of the ground state of the rotor model
for different boundary conditions. It turned out that these components could
be normalised in such a way that they are all positive integers. Moreover,
taking the greater common divisor equal to $1$, their sum was given by
an enumeration of certain classes of ASMs with extra symmetry.


Motivated by these results, we study the rotor model with spectral
parameters. As usual the value of the degree of the minimal polynomial
solution is quite hard to derive. Here we make the assumption, justified by
the explicit solutions for small sizes and \emph{a posteriori} by the
evaluation at the homogeneous point, that the degree is two times the
degree of the corresponding fully packed $O(1)$ model.  

Then we find on one hand that the polynomial sum rules are quite easy
to derive, using a mapping to the $O(1)$ model. 
On the other hand contrarily to the $O(1)$ model, where the smallest component
is given by a product of degree-one 
terms, for the rotor model there are no completely factorized
components. Nonetheless we are able to compute explicitly a whole family
of terms which we call \emph{maximally nested components} and which in
most cases contains the smallest component. We see in this context the appearance
of $3-$enumerations of alternating sign matrices.

The plan of the article is as follows: in Sect.~\ref{description} we
give a brief description of the rotor model with different boundary
conditions. In Sect.~\ref{spectral} we 
introduce the spectral parameters. We find a mapping of our model to the
fully packed $O$(1) model, which will allow us to derive most of the
recursion relations we need. We write also the exchange equation.
Then in Sect.~\ref{sumsMNC} we come to the main results of
our paper: the sum rules and the maximally nested components for periodic
and open boundary conditions. 

\section{\bf The rotor model}\label{description}

The rotor model \cite{martins-1998, rotor} is defined on a square lattice. 
In the interior of a face there are
four lines that can be thought as the paths
followed by four cars coming from different directions at a crossroad,
under the condition that each car 
turns right or left and no two cars go in the same direction. 
Under this restriction there are only the four possible
configurations for a face 
called $R, L, A, D $. Each configuration appears with
probability respectively $\omega_R, \omega_L, \omega_A, \omega_D$.

\begin{center}
\includegraphics{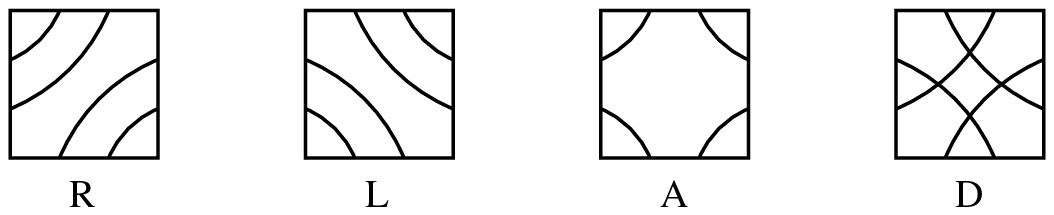}

Configurations.
\end{center}

\noindent

Once all the faces of the lattice are assigned, one gets a global
configuration consisting of closed 
loops and, in presence of boundaries, open curves connecting points on
the boundary.  
We are interested in geometries where the lattice is semi-infinite in
one direction (the vertical one) and finite in the other.
Let us label the rows ascending from 1 to $\infty$
and the columns from left to right
in such a way that each face is labelled by a pair $(i,j)$.
Now, depending on the parity of $i+j$, let us colour the lines of our
configurations in red and green in the following way
\sk
\begin{center}
\includegraphics{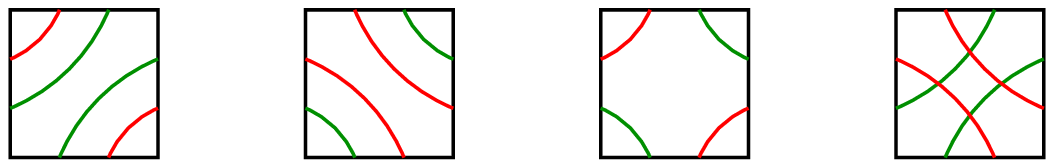}

Coloration for $i+j$ even.
\end{center}

\begin{center}
\includegraphics{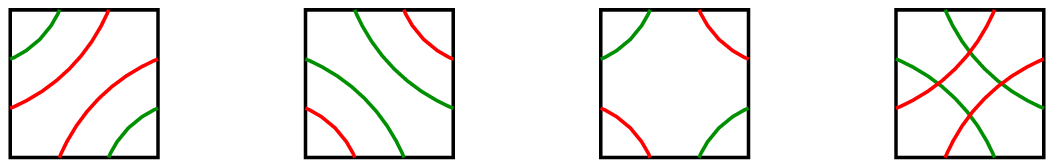}

Coloration for $i+j$ odd.
\end{center}

\noindent

\sk
\begin{figure}[h]
\centering
\includegraphics{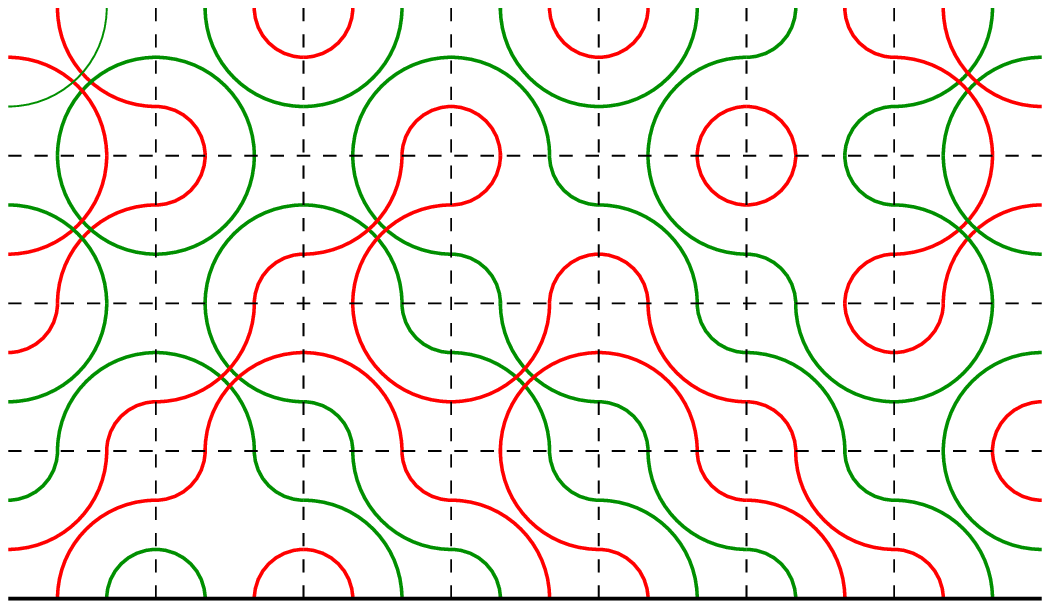}
\caption{An example of the result of the
coloration on a sample configuration}\label{Example1}
\end{figure}

\begin{figure}[h]
\centering
\includegraphics{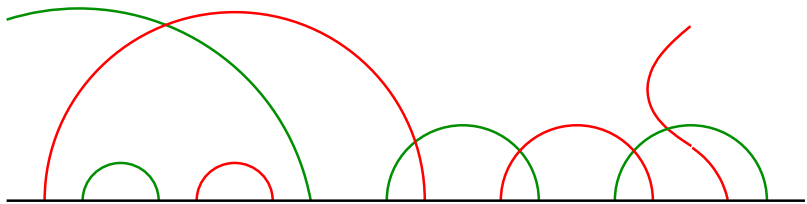}
\caption{Link pattern corresponding to the sample of
  fig.~\ref{Example1}}\label{Example2} 
\end{figure}

As one can see from fig.~\ref{Example1} red (green) lines join only
red (green) lines and lines of different colours can cross (inside a $D$
face), while lines of the same colour cannot.
The points on the boundary can be connected only if
they have the same colour, and looking at each colour separately we will
have a link pattern of non-crossing lines like in the fully packed
$O(1)$ case. 
For example, the connectivities of fig.~\ref{Example1} give the
link pattern of fig.~\ref{Example2}
\sk

In the horizontal direction we always choose
boundary conditions which respect 
the colouration. They can be periodic (PBC) or closed (CBC). 
The periodic boundary conditions for lattice of even size of course preserve
the coloration.  
In the case of a lattice of odd size we have to introduce 
a vertical seam that exchanges the colours in the horizontal direction.
We are interested in the connectivities, hence each configuration
corresponds to a pair of non-intersecting link patterns (red and
green) which for the 
periodic lattice can be draw a disk. Actually for even period
lattices we could also
take trace of the presence of the hole in the cylinder, in this case
we call the boundary conditions PBC+$\infty$ and each configuration is
mapped to two link patterns on a  punctured disk.

The closed boundary conditions are defined in the following way: take the
four points on the 
right 
boundary of a pair of rows $2i-1, 2i$ and connect them in a way consistent
with the colouration 
(do the same on the left). Here the connectivities can be drawn on the
upper-half plane.

\begin{center}
\includegraphics{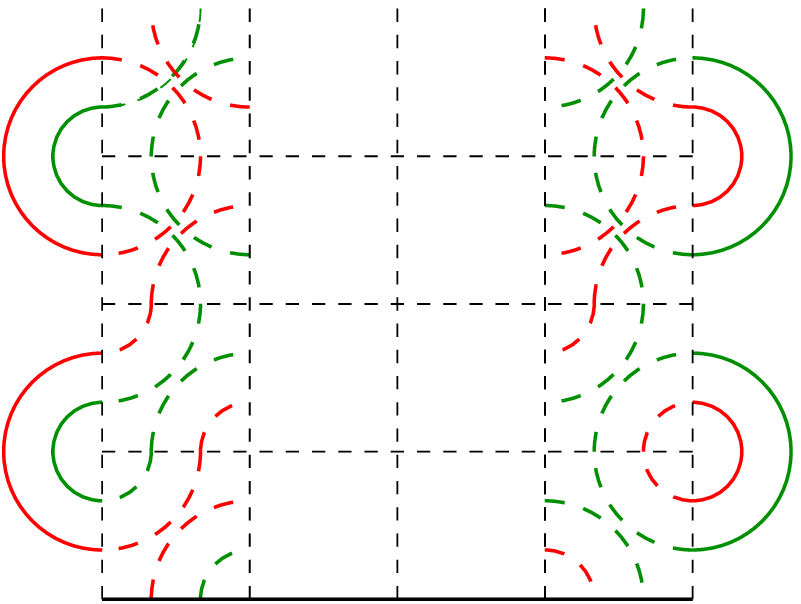}

Closed boundary conditions.
\end{center}

We are interested in the probability
of having a certain connectivity of 
the base points or, in the light of what we have said above,
of having a pair of link patterns $(\pi_G, \pi_R)$. 

The standard technique to deal with this kind of problems makes use of 
the transfer matrix $T$, whose effect is that of adding one (or two for CBC)
more rows to the 
cylinder. The link patterns probability has to be stationary
under this action. This simply means that it forms an
eigenvector of $T$ with eigenvalue 1.

Having given the general setting, we restrict our attention to the
choice of the weights that gives an integrable system. 
Let us introduce the $R-$matrix
\begin{center}
\includegraphics{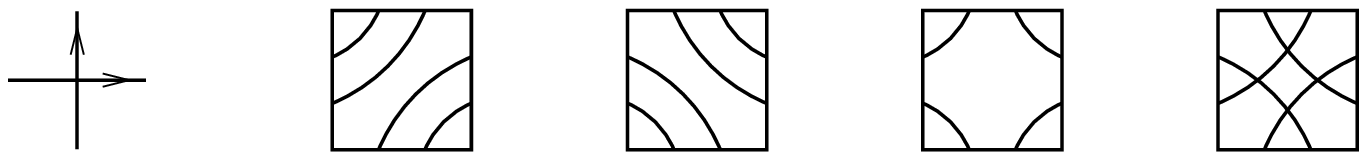}
\put(-345,18){$=$}
\put(-320,18){$\omega_A$}
\put(-250,18){$+$}
\put(-235,18){$\omega_D$}
\put(-165,18){$+$}
\put(-150,18){$\omega_L$}
\put(-80,18){$+$}
\put(-65,18){$\omega_R$}
\end{center}

\noindent
in terms of which the one-row transfer matrix for periodic lattices reads
\sk
\begin{center}
\includegraphics{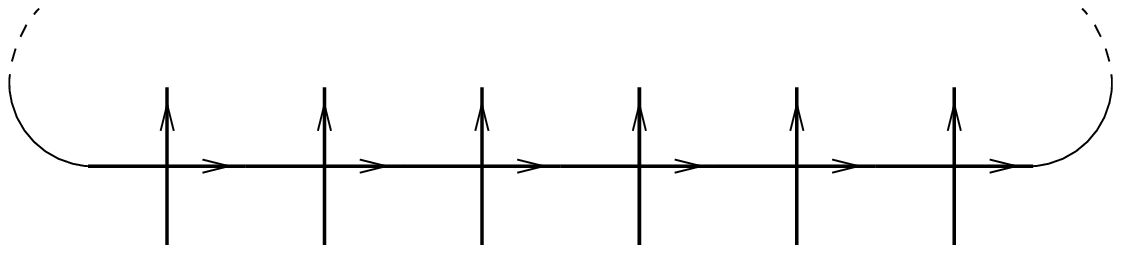}
\end{center}

For CBC the double-row transfer matrix is given in terms of the
$R-$matrix introduced above and the boundary $K-$matrix, which in our case has
the simple effect of inverting the spectral parameter

\sk

In \cite{martins-1998} Martins and Nienhius have solved the
Yang-Baxter equation for the 
previous R-matrix, finding three classes of solutions. Here, following
\cite{rotor},  we
consider only what in \cite{martins-1998} is  called class II
\begin{gather}\label{def_RLAD}
\omega_R(z,w) = \omega_L(z,w) ~=~ (w-z)(z+qw);\\
\omega_D(z,w) ~=~ q(w-qz)(z+qw)\\
\omega_A(z,w) ~=~ (w-z)(w+qz)
\end{gather}
where $q = e^{2\pi i/3}$ and $z, w$ are spectral parameters.

Notice that we have \emph{not} normalised the sum of the $\omega_i$ to be
$1$, but
$
\omega_R + \omega_L + \omega_A + \omega_D = qz^2 - w^2
$

What Batchelor et al. have done in
\cite{rotor} is to compute the
probabilities of link-patterns for different boundary
conditions and different lattice size. In terms of the eigenvector of
the transfer matrix they have found that one can choose a
normalisation such that all the components are integers and the GCD
equals 1. 
In particular they were able to identify the sum $S$ of the components
and in case of a odd size lattice with CBC the smallest component:

\begin{itemize}
\item
{\bf PBC} -\emph{ Even}:
$S(2n)= 3^{\theta_n} A(n;1)$;
\item
{\bf PBC} -\emph{ Odd}:
$S(2n+1) = 3^{3n} A_V(2n-1;3)^2$;
\item
{\bf PBC+$\infty$} -:~
$S(2n) = 3^{n^2} A_{HT}(2n)$;
\item
{\bf CBC} -\emph{ Even}: $S(2n) = 3^{2\theta_n} A_V(2n + 1)$;
\item
{\bf CBC} -\emph{ Odd}: 
$S(2n-1) = 3^{(n-1)^2} \sN_8(2n)$;
\end{itemize}
where $\theta_n = \lfloor(n-1)(n+2)/3 \rfloor$ and the other quantities will
be defined in the following.

In the present paper we not only prove these conjectures but we also
determine the value of a family of components having maximally nested
arcs. This family, in all cases except for CBC and even size, contains the
smallest component. 


In the case of lattices of size $2n$ with PBC, each MNC is given by the
product of two 3-enumerations of ASMs $A(m;3)A(k;3)$, where $m$ and $k$
represent the relative orientation of the red and green link patterns. The
smallest component is then given by $A(\lfloor n/2 \rfloor;3)A(\lfloor (n+1)/2
\rfloor;3)$. 

For periodic systems in which we keep track of the hole in the cylinder we
give a Pfaffian formula for the values of the MNCs, but we have not been
able to recognise these numbers as enumerations of some known objects. 

In the case of closed boundary conditions, we have computed the
smallest component for odd size, and found $A_V(2n+1;3)$ (this in fact is
the only value predicted in \cite{rotor}). For lattices of even size we have
computed the MNC with all arcs parallel (which is not the smallest one), and
found again $A_V(2n+1;3)$.

\section{Spectral parameters}\label{spectral}

Up to now the probabilities $\omega_R, \omega_L, \omega_A, \omega_D$
were independent of the face position. The key idea of Di Francesco
and Zinn-Justin \cite{pdf-pzj-1} was to generalise the problem
considering different 
spectral parameters $z_i$ for each vertical line. In the following we
will explicitly treat the case of PBC and even size lattice, the other
cases being completely analogous. At the proper moment we will point out the
differences. 

In presence of spectral parameters the transfer matrix $T(t|z_1,
\dots, z_{2n})$ becomes 
\eq\label{eigenvector}
T(t|z_1, \dots, z_{2n}) = {\rm Tr}R(z_1, t)R(z_2, t)\dots R(z_{2n}, t)
\en

Because of the normalisation, 
the eigenvector equation we
have to study assumes the form 
\eq
T(t|z_1, \dots, z_{2n}) \Psi(z_1, \dots, z_{2n}) =
\prod_{i=1}^{2n}(qz_i^2-t^2)~ \Psi(z_1, \dots, z_{2n}) 
\en
Introducing a basis of of pairs of link patterns (red and green) $|\pi_R\>>
\otimes|\pi_G\>>$, we write 
\eq
\Psi(z_1, \dots, z_{2n}) = \sum_{\pi_R, \pi_G} \Psi_{\pi_R, \pi_G}(z_1, \dots,
z_{2n}) |\pi_R\>> \otimes|\pi_G\>>
\en
where, if we normalise the sum to be $1$, $\Psi_{\pi_R, \pi_G}(z_1, \dots,
z_{2n})$ is the probability of having the link pattern configuration
$\pi_R, \pi_G$. 

A couple of remarks are in order. 
\begin{itemize}
\item[{\bf 1.}] The fact that $\Psi$ doesn't depend on $t$ is a
   trivial consequence 
   of the commutativity of the transfer matrices for different values
   of the parameter $t$. 
$$
[T(t|z_1, \dots, z_{2n}),T(t'|z_1, \dots, z_{2n})] = 0.
$$

\item[{\bf 2.}] It is easy to see that the transpose of the transfer
   matrix has a 
   very simple eigenvector of eigenvalue $\Lambda =
   \prod_{i=1}^{2n}(qz_i^2-t^2)$, namely
   \eq\label{left-eigen}
   \<< \Omega| := \left(\sum_{\pi_R} \<<\pi_R|\right) \otimes
   \left(\sum_{\pi_G} \<<\pi_G|\right). 
   \en
   It is simply the functional which gives 1 on each link configuration.

\item[{\bf 3.}] The solution of eq.(\ref{eigenvector}) can be normalised to be a
	 homogeneous polynomial. We require it to be of \emph {minimum degree}.

\end{itemize}
\sk
Let us introduce the matrix $\check{R}$
\eq
\check{R}_i(z, w)= q(w - q z)(q w + z) {\rm Id} +  (w - z)(w + qz) E_i +
(w-z)(q w + z) (R_i + L_i)
\en

The operators $E_i, R_i, L_i$ for $i=1,\dots, N$ satisfy the following
relations 
\eq
\begin{split}
&E_i = R_i L_i = L_i R_i,~~~~~~R_i^2 = R_i ~~~~~~ L_i^2= L_i\\
&L_i R_{i\pm 1} L_i = L_i, ~~~~~~ R_i L_{i\pm 1} R_i = L_i, ~~~~~~[R_i,
R_{i\pm 1}] =[L_i, L_{i \pm 1}] =0\\
&[R_i, R_j] =[L_i, L_j] =[R_i, L_j]=0~~~~|i-j| \geq 2
\end{split}
\en
and  have the following graphical representation
\begin{center}
\includegraphics{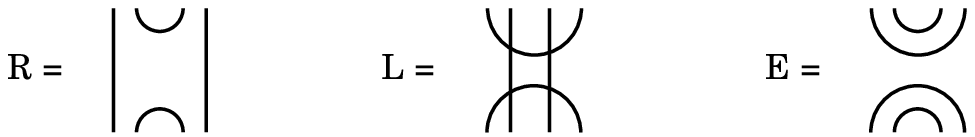}
\end{center}

The $R-$matrix/$\check{R}-$matrix and the $K-$matrix, besides the
Yang-Baxter and the boundary Yang-Baxter equation,
satisfy the unitarity property
\sk
\begin{center}
\includegraphics{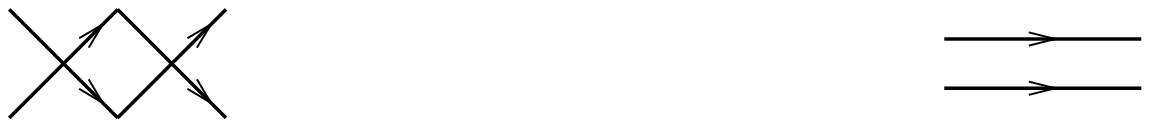}
\put(-235,12){$=~~ q^2 (w^2 - q^2z^2)(z^2 - q^2w^2)$}
\put(-340,-3){$w$}
\put(-340,27){$z$}
\put(-70,0){$w$}
\put(-70,24){$z$}
\end{center}

\noindent
and the inversion relation
\sk
\sk
\begin{center}
\includegraphics{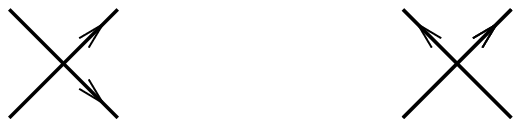}
\put(-160,-3){$w$}
\put(-160,27){$z$}
\put(-45,4){$w$}
\put(-2,4){$-q^2 z$}
\put(-100,12){$= ~~ -q$}
\end{center}
 
\subsection{Mapping to the $O(1)$ model
}\label{mapping}

As described above, the vector space of configurations of the rotor
model can be seen as a tensor product of two copies of the vector space of
the $O(1)$ model. Tracing on one of the two colours (say tracing on
green) is equivalent to mapping the rotor model on the $O(1)$
model. It is very easy to realise the effect of the trace $\sT = \sum
\<< \pi_G|$ on the transfer matrix or on the $\check{R}$ matrix 
\eq\label{tracing1}
\sT T(t|z_1, \dots, z_{2n}) = T_{O(1)}(t^2|z_1^2, \dots, z_{2n}^2) \sT,
~~~~~~
\sT\check{R}(z_i, z_{j}) = \check{R}_{O(1)}(z_i^2, z_{j}^2) \sT.
\en
In particular this means that for the eigenvectors is valid the following
\eq\label{Psi_prop_Psi}
\sT\Psi(z_1,\dots, z_{2n}) \propto \Psi_{O(1)}(z_1^2,\dots,
z_{2n}^2). 
\en
In this paper we make the assumption that the degree of
$\Psi$ in each variable is two times the degree of
$\Psi_{O(1)}$. Hence we can choose the normalisation of $\Psi$ in such
a way  that the proportionality in eq.(\ref{Psi_prop_Psi}) becomes an 
equality.

\subsection{Exchange equations}\label{qKZsection}

As a consequence of the Yang-Baxter equation we have the following
\begin{proposition}\label{commutation}
The transfer matrices $T(t;\dots, z_i, z_{i+1},\dots)$ and
$T(t;\dots, z_{i+1}, z_{i},\dots)$ are intertwined  by
$\check{R}_{i,i+1}(z_i, z_{i+1})$, i.e. 
\eq  
T(t;.., z_i, z_{i+1},..) \check{R}_{i,i+1}(z_i,
z_{i+1})  = \check{R}_{i,i+1}(z_i, z_{i+1})T(t;.., z_{i+1},
z_{i},..). 
\en 
\end{proposition} 
\begin{center}
\includegraphics{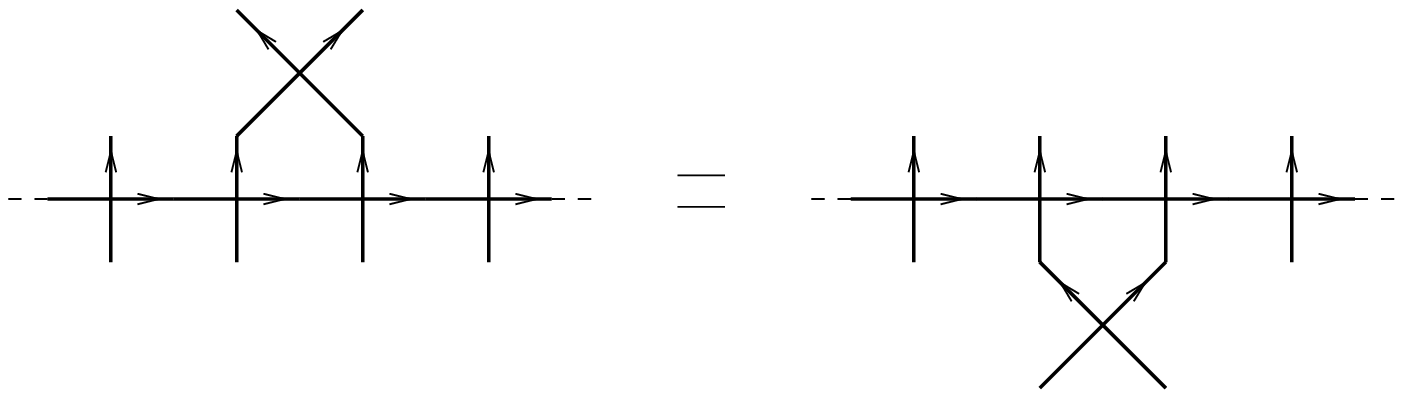}
\put(-110,-7){$z_i$}
\put(-73,-7){$z_{i+1}$}
\put(-73,80){$z_i$}
\put(-110,80){$z_{i+1}$}
\put(-345,28){$z_i$}
\put(-310,28){$z_{i+1}$}
\end{center}

From proposition (\ref{commutation}) it follows 
\eq\label{qKZ} 
\check{R}_{i,i+1}(z_i, z_{i+1}) \Psi(.., z_{i+1}, z_{i},..) = \alpha_i
(z_1,..,z_{i}, z_{i+1},.., z_{2n}) \Psi(.., z_{i}, z_{i+1},..) 
\en 
where $\alpha_i$ is a polynomial which can be determined using the
projection to the $O(1)$ model
$$
\alpha_i(.., z_i, z_{i+1},..) = \alpha(z_i, z_{i+1}) = (q z_{i+1}^2 - z_i^2). 
$$
Alternatively, for even PBC one can derive $\alpha_i$ without making
use of the projection. From  the unitarity condition it follows 
\eq\label{alpha-alpha} 
\alpha_i (z_1,.., z_{i}, z_{i+1},.., z_{2n})\alpha_i (z_1,.., z_{i+1},
z_{i},.., z_{2n}) = q^2 (z_i^2 - q^2z_{i+1}^2)(z_{i+1}^2 - q^2z^2_i), 
\en 
then we notice that the left eigenvalue defined in
eq.(\ref{left-eigen}) satisfies 
\eq\label{projXR} 
\<<\Omega| \check{R}_{i,i+1}(z_i, z_{i+1}) = q(z_{i+1} + q z_i)(z_{i+1} - q
z_i)\<<\Omega| 
\en
Hence 
\eq
\begin{split}
\prod_{i=1}^{2n-1} (q z_{2n}^2 - z_i^2)~~ &\<<\Omega| \Psi(z_{2n}, z_{1},
.., z_{2n-1})\>> \\ 
=\<<\Omega| \check{R}_{2n-1}(z_{2n}, z_{2n-1})\dots \check{R}_{2}(z_{2n},
z_{2})&\check{R}_{1}(z_{2n}, z_{1}) \Psi(z_{2n}, z_{1}, .., z_{2n-1})\>>\\  
=\left(\prod_{i=1}^{2n-1} \tilde\alpha_i \right) ~~&\<<\Omega|\Psi(z_{1},
z_{2}, .., z_{2n})\>>.\\ 
\end{split} 
\en 
Where $\tilde\alpha_i = \alpha(z_1,..,z_{i-1}, z_{2n},
z_{i+1},..,z_{2n-1},z_i)$. 
But the invariance of the system under discrete translation in the
horizontal direction tells us that $\<<\Omega|\Psi(z_{1}, z_{2}, ..,
z_{2n})\>> = \<<\Omega| \Psi(z_{2n}, z_{1}, .., z_{2n-1})\>>$, then 
$
\prod_{i=1}^{2n-1} (qz_{2n}^2 - z_i^2) = \prod_{i=1}^{2n-1} 
\tilde\alpha_i   
$
. This, combined with eq.(\ref{alpha-alpha}) 
fixes $\alpha_i(z_1,..,z_{2n}) = q(z_{i+1} + q z_i)(z_{i+1} - q z_i)$.

Eq.(\ref{qKZ}) will be one of our main tools to analyse $\Psi(z_1,..,
z_{i}, z_{i+1},.., z_{2n})$.
Let us write it in components. 
\begin{itemize}
\item First let $\Psi_{\pi_0, \pi'_0}$ be a component
which has no little arcs connecting $i$ and $i+1$, then
\eq
q(z_{i+1}- q z_i)(q z_{i}+ z_{i+1}) \Psi_{\pi_0, \pi'_0}(.., z_{i}, z_{i+1}, ..) =
q(z_{i+1}- q z_i)(q z_{i+1}+ z_i) \Psi_{\pi_0, \pi'_0}(.., z_{i+1}, z_i, ..) 
\en
This means that $\Psi_{\pi_0, \pi'_0}(.., z_i, z_{i+1}, ..) =
(z_{i} + q z_{i+1}) \tilde\Psi_{\pi_0, \pi'_0}(.., z_i, z_{i+1}, ..) $, 
where \\
$\tilde\Psi_{\pi_0, \pi'_0}(.., z_i, z_{i+1}, ..)$ is symmetric 
under exchange of $z_i$ and $z_{i+1}$.

In general if we consider a component with 
no arcs (of any colour) in between $i$ and $j$ 
(with respect to the ordering), then we can write
\eq\label{symmetric1}
\Psi_{\pi_0, \pi'_0}(.., z_i, .., z_{j}, ..) = 
\prod_{i \leq l < m \leq j} (z_{l} + q z_{m})\tilde\Psi_{\pi_0, \pi'_0}
(.., z_i, .., z_{j}, ..),  
\en
where $\tilde\Psi_{\pi_0, \pi'_0}(.., z_i, .., z_{j}, ..)$ is symmetric 
under exchange of $z_l$ and $z_m$ with $i \leq l < m \leq j$.

\item Let now $\Psi_{\pi_c, \pi'_0}$ be a component
which has a little red arc connecting $i$ and $i+1$ and no green arc, then
\eq\label{exch2}
\begin{split}
q(z_{i+1}- q z_i)(z_{i+1}+ q z_{i}) \Psi_{\pi_c, \pi'_0}(.., z_{i}, z_{i+1},
..)\\ =   q(z_i + q z_{i+1})(z_i - q z_{i+1})\Psi_{\pi_c, \pi'_0}(.., z_{i+1},
z_{i}, ..)\\ + \sum_{(\pi_0,\pi'_0)} (z_{i+1} - z_i)(q 
z_i + z_{i+1}) \Psi_{\pi_0, \pi'_0}(.., z_{i+1}, z_{i}, ..)
\end{split}
\en
where the sum is over all the diagrams that have no arcs connecting $i$ and
$i+1$ and are mapped to the pair $(\pi_c, \pi'_0)$ under the action of $R_i$
or $L_i$. 

\item If $\Psi_{\pi_c, \pi'_c}$ is a component
which has both a red and a green arcs connecting $i$ and $i+1$, then
\eq
\begin{split}
q(z_{i+1} - q z_i)(z_{i+1}+ q z_{i}) \left(\Psi_{\pi_c, \pi'_c}(.., z_{i}, z_{i+1},
..) - \Psi_{\pi_c, \pi'_c}(.., z_{i+1},
z_{i}, ..)\right) \\ = q^2(z_i^2 - z_{i+1}^2) \left(\sum_{(\pi_c, \pi_o)}
\Psi_{\pi_c, \pi'_o}(.., 
  z_{i+1}, z_{i}, ..) + \sum_{(\pi_o, \pi_c)}  \Psi_{\pi_o, \pi'_c}(..,
  z_{i+1}, z_{i}, 
  ..)\right) \\
+ (z_{i+1} - z_{i})(z_i+qz_{i+1})\sum_{(\pi_o, \pi'_o)}  \Psi_{\pi_o, \pi'_o}(.., z_{i+1}, z_{i},
  ..).
\end{split}
\en
The first and second sums are over diagrams which have a single arc
connecting $i$ and $i+1$ and are mapped to $(\pi_c, \pi'_c)$ by $E_i$ and $R_i$
or $L_i$. The third sum is over diagrams which have no arcs
connecting $i$ and $i+1$ and are mapped to $(\pi_c, \pi'_c)$ by $E_i$.
\end{itemize}

\noindent
We see from the previous remarks that for $z_{i+1} = -~q^2 z_i$ we have
\eq
\check{R}_i(z_i, -q^2 z_i) = (q^2 -1)z_i^2 E_i~~,
\en
hence
\eq\label{proj2}
\Psi(.., z_{i}, -q^2 z_{i}, ..) = E_i \Psi(.., -q^2 z_{i}, z_{i},
..). 
\en
This means that $\Psi_{\pi_R, \pi_G}(.., z_i, -q^2 z_i, ..)=0$ 
whenever both $\pi_R$ and $\pi_G$ have no arcs connecting points $i$ 
and $i+1$. 

While for $z_{i+1} = q z_i$
\eq
\check{R}_i(z_i, q z_i) =  (q^2 - q)(2 E_i - R_i - L_i)  ~~,
\en
hence
\eq
(2 E_i - R_i - L_i)\Psi(..,  z_{i}, q z_{i}, ..) = 0.
\en
Since $(E_i-R_i)$ and $(E_i - L_i)$ are orthogonal projectors we have
both 
\eq\label{proj1}
(E_i - R_i)\Psi(..,  z_{i}, q z_{i}, ..) = 0 ~~~~~~{\rm and}  ~~~~~~
(E_i - L_i)\Psi(..,  z_{i}, q z_{i}, ..) = 0. 
\en

When written in components, eq.(\ref{proj1}) implies that for $z_{i+1} 
= q z_i$ looking at the configurations whose points $i$ and $i+1$ are 
\emph{not} connected by a red arc one finds far all green $\pi_G$ 
\eq\label{zero2}
\sum_{\substack{\pi'_G\\e_i \pi'_G ~= ~e_i \pi_G}}
\Psi_{\pi_R,\pi'_G}(..,  z_{i}, q 
z_{i}, ..) = 0, 
\en
Where $e_i$ is the usual Temperley-Lieb generator acting on link patterns. 
An analogous statement is of course valid for red and green exchanged.

\subsection{Recursion relations}

We want to write now a recursion relation for the components 
of the ground state.  
This is possible because the vector space of configurations of the rotor
model on a lattice of size $N$ can be mapped to the vector space of
configurations of size $N+2$ in a trivial way, simply adding both a
green and a red arc in between points $i-1$ and $i$.

Let us call this map $\phi_i$, then as a simple consequence of
the unitarity and the inversion relation, we have the following
\eq\label{Tphi}
T_N(z_1, .., z_i, z_{i+1} = -q^2 z_i, .., z_{N}) \phi_i ~
  \propto ~\phi_i T_{N-2}(z_1, ..,\hat{z}_i,\hat{z}_{i+1} .., z_{N}). 
\en
Using eq.(\ref{Tphi}) and eq.(\ref{proj2}) we conclude that
\eq\label{propor1}
\Psi^{(N)}(z_1, .., z_i, z_{i+1} = -q^2 z_i, .., z_{N}) \propto \phi_i
\Psi^{(N-2)}(z_1, .., \hat{z}_i,\hat{z}_{i+1}, .., z_{N}). 
\en
\begin{center}
\includegraphics{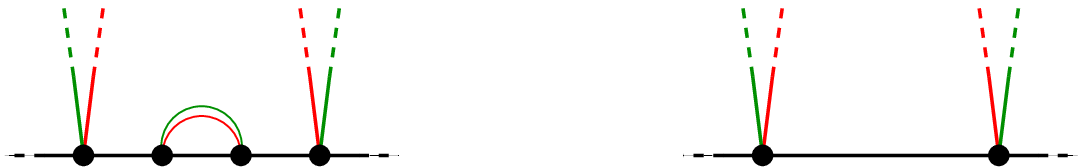}
\put(-30,-9){$z_{i+2}$}
\put(-95,-9){$z_{i-1}$}
\put(-245,-9){$z_{i+1}= -q^2 z_i$}
\put(-272,-9){$z_i$}
\put(-295,-9){$z_{i-1}$}
\put(-160,15){\large{$\propto$}}
\end{center}
Both eq.(\ref{Tphi}) and eq.(\ref{propor1}) are of course valid
independently of the specific boundary conditions we choose. What
changes is the proportionality  factor in eq.(\ref{propor1}), which
can be derived using the mapping
to the $O(1)$ model given in section \ref{spectral}. For the periodic
even lattice it reads 
\eq
\begin{split}
\sT~\Psi^{(2n)}(z_1, .., z_i, z_{i+1} = -q^2 z_i, .., z_{2n}) = 
\Psi^{(2n)}_{O(1)}(z_1^2, .., z_i^2, z_{i+1}^2 = q z_i^2, .., z_{2n}^2) \\ 
= \prod_{j\neq i,i+1}q(z_{i+1}^2 - q^2 z_j^2)
\phi_i\Psi^{2(n-1)}_{O(1)}(z_1^2, .., \hat{z}_i,\hat{z}_{i+1}, .., z_{2n}^2) \\
=\prod_{j\neq i,i+1}q(z_{i+1}^2 - q^2 z_j^2)
\sT\phi_i\Psi^{2(n-1)}(z_1, .., \hat{z}_i,\hat{z}_{i+1}, .., z_{2n}), 
\end{split}
\en
where $\phi_i$ in the second line is intended to act on configurations of
the $O(1)$ model, adding an arc joining $i-1$ and $i$.
Actually for the periodic even lattice we can recover the previous result
without recurring to the projection; in facts 
we can argue that if we set $z_{i+1} = -q^2
z_i$ and $z_{i+2} = -q^2 z_{i+1}$ we must have $\Psi^{(2n)}(z_{i+1} = -q^2
z_i)$ in the image of 
both $E_i$ and $E_{i+1}$. This is compatible only with 
$\Psi^{(2n)}(z_1, ..,
z_i, z_{i+1} = -q^2 z_i, z_{i+2} = q z_i, .., z_{2n}) = 0.
$ Using repeatedly eq.(\ref{qKZ}) we find 
$\Psi^{(2n)}(z_1, ..,
z_i, z_{i+1} = -q^2 z_i, .., z_{j} = q z_i, .., z_{2n}) = 0
$. If instead we set $z_{i+1} = -q^2 z_i$ and $z_{i+2} = q z_{i+1}$ we
must have $\Psi^{(2n)}(z_1, .., 
z_i, z_{i+1} = -q^2 z_i, z_{i+2} = - z_i, .., z_{2n})$ in the image of
$E_i$ and in the kernel of both $(E_{i+1} - R_{i+1})$ and $(E_{i+1} -
L_{i+1})$. Again this is compatible only with $\Psi =0$ and using
repeatedly eq.(\ref{qKZ}) we get
$\Psi^{(2n)}(z_1, ..,
z_i, z_{i+1} = -q^2 z_i, .., z_{j} = - z_i, .., z_{2n}) = 0
$. Coming back to eq.(\ref{propor1}) what we have found implies 
\eq\label{propor2}
\Psi^{(2n)}(z_1, .., z_i, z_{i+1} = -q^2 z_i, .., z_{2n}) = 
\kappa \prod_{j\neq i,i+1}q(z_{i+1}^2 - q^2 z_j^2) \phi_i 
\Psi^{(2n-2)}(z_1, .., \hat{z}_i,\hat{z}_{i+1}, .., z_{2n}). 
\en
Since the degrees of both
sides match,  $\kappa$ is a pure number that we can fix to $1$ 
We have proved the following recursion relation
\begin{equation}\label{propor3}
\Psi^{(2n)}(..,z_i, z_{i+1} = -q^2 z_i, ..) = 
\prod_{j\neq i,i+1}q(z_{i+1}^2 - q^2 z_j^2) 
\phi_i
\Psi^{(2n-2)}(.., \hat{z}_i,\hat{z}_{i+1}, ..). 
\end{equation}

We can now pass to the computation of the sum of the components and of the
of the maximally nested components.

\section{Sums and Maximally Nested Components}\label{sumsMNC}

\subsection{PBC even}


\subsubsection*{Sum of the components}

Let us denote the sum of the components of $\Psi^{(2n)}$ by 
$\somma_{2n}(z_1, .., z_{2n}) 
= 
\<<\Omega_{2n}|\Psi^{(2n)}(z_1, .., z_{2n})\>> 
$. It 
can be easily derived using the projection to the $O(1)$ model \cite{pdf-pzj-1} 
\eq\label{sum=}
\somma_{2n}(z_1, z_2, .., z_{2n})  =
\somma^{O(1)}_{2n}(z_1^2, z_2^2, .., 
z_{2n}^2) = S_{Y_n}(z_1^2, z_2^2, .., z_{2n}^2),
\en 
where $S_{Y_n}(z_1, z_2, .., z_{2n})$ is the Schur function corresponding to
the Young diagram $Y_n$ with two rows of length $n-1$, two rows of length
$n-2$, .., two rows of length $2$ and two rows of length $1$.

Notice that the first equality in eq.(\ref{sum=}) remains valid for every
boundary condition. 
Specialising eq.(\ref{sum=}) to $z_i= 1$ we obtain \cite{pdf-pzj-1,
okada, strog1}
\eq\label{sum=1}
\begin{split}
\somma_{2n}(1, .., 1) = S_{Y_n}(1, .., 1) =
3^{\frac{n(n-1)}{2}}A(n;1) =~~~~~~~~~~~~~~~\\
3^{\frac{n(n-1)}{2}}
\prod_{i=1}^n \frac{(3i-2)!}{(n+i-1)!} = 1,~~ 3\cdot 2,~~ 3^3\cdot 7,~~
3^6\cdot 42,~~ 3^{10}\cdot 429,~~ 3^{15} \cdot 7436, ...
\end{split}
\en
$A(n;1) $ is the enumeration of alternating sign matrices of size $n$.
\sk

\subsubsection*{The maximally nested components}

The recursion relation in eq.(\ref{propor3}) allows us to derive explicitly
the components
corresponding to the maximally nested configurations.
Let us first consider the parallel one,   
which has arcs of both colours connecting the pairs $(i, 2n-i+1)$. 
\begin{center}
\includegraphics{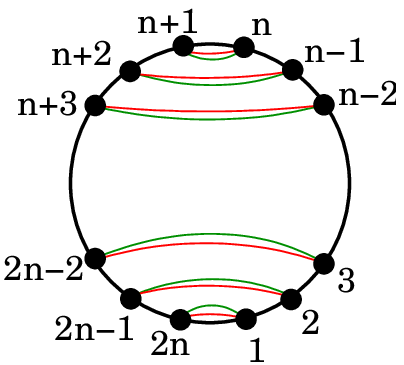}
\end{center}
This configuration has no arcs of any colour in between points $1$ and $n$
and in between points $n+1$ and $2n$. Hence, 
calling the corresponding component $\Psi_{0,n}$, 
from eq.(\ref{symmetric1}) we can write
\eq\label{Psi-tilde}
\Psi_{0,n}(z_1, .., z_{2n}) =\!
\prod_{1 \leq i < j \leq n} \!(qz_j + z_i)\!\!\!
\prod_{n+1 \leq i < j \leq 2n} \!\!\!(-q^2z_j - q z_i)
~\tilde\Psi_{0,n}(z_1, .., z_{2n}),
\en
where $\tilde\Psi_{0,n}$ is a homogeneous polynomial 
of total degree $n(n-1)$, $n-1$ in each variable and symmetric  
separately in $z_1, \dots, z_n$ and $z_{n+1}, \dots, z_{2n}$.

Combining the definition of $\tilde\Psi_{0, n}$ 
from eq.(\ref{Psi-tilde}) with eq.(\ref{propor3}), 
we find the following recursion relation for $\tilde\Psi_{0, n}$
\eq\label{recur-tilde}
\tilde\Psi_{0, n}(z_1, .., z_{2n} = -q z_1) =
\prod_{j=2}^n q (z_1 - q z_j)
\prod_{j=n+1}^{2n-1} q (z_1 + q z_j)
\tilde\Psi_{0, n-1}(z_2, .., z_{2n-1})
\en
The recursion relation eq.(\ref{recur-tilde}) has a unique solution 
with initial condition $\tilde\Psi_{0,1}=1$, degree $n-1$ 
in each variable and with the same symmetries of $\tilde\Psi_{0, n}$, namely
\eq\label{sol-Psi-tilde}
\tilde\Psi_{0, n}(z_1, .., z_n, z_{n+1}, .., z_{2n}) = 
q^{2n(n-1)} S_{Y_n} (z_1, .., z_n, -z_{n+1}, .., -z_{2n}),
\en
hence for $\Psi_{0,n}$ we get
\eq\label{sol-Psi}
\Psi_{0, n} = 
q^{2n(n-1)} 
\prod_{1 \leq i < j \leq n} (qz_j + z_i)\!\!\!
\prod_{n+1 \leq i < j \leq 2n} \!\!\!(-q^2z_j - q z_i)~
S_{Y_n} (z_1, .., z_n, -z_{n+1}, .., -z_{2n}).
\en
In particular if we specialise to all $z_i=1$ we obtain
\eq\label{sol-Psi1}
\Psi_{0, n}(1, .., 1, .., 1) = 
(-1)^{\frac{n(n-1)}{2}} 
S_n (1, .., 1, -1, .., -1).
\en

In order to recognise the right-hand side of eq.(\ref{sol-Psi1}), we recur
to the well known one-to-one correspondence between ASM's and $6-$vertex
configurations with domain wall boundary conditions.
If the weights of the vertex configurations are given by
\sk
\begin{center}
\includegraphics{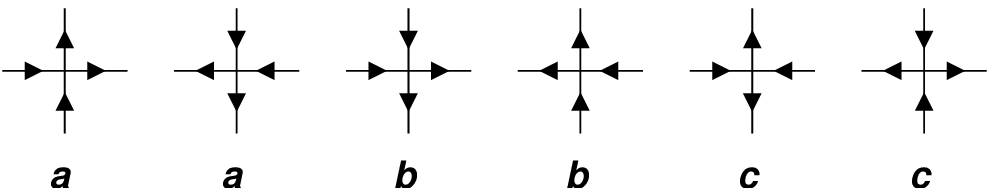}
\end{center}
with $a = q x - q^2 y  ,~ b =  q y - q^2 x ,~ c= (q^2 - q)(x y)^{1/2} $
and $q = e^{2\pi i/3}$, then the partition function of the model is
given by \cite{izergin, okada, strog1}
\eq
Z_n(q = e^{\frac{2\pi i}{3}}| x_1, .., x_n; y_1, .., y_n) 
= (-1)^{\frac{n(n-1)}{2}}(q^2-q)^n
\prod_{j=1}^n x_j^{1/2}\prod_{j=1}^n y_j^{1/2}
S_{Y_n}(x_1, .., x_n; y_1, .., y_n).
\en
If we now set $x_1 = .. = x_n = 1$ and $y_1 =  ..= y_n = -1$ we see that
configurations (a) and (b) have weight $\pm 1$ (the sign actually doesn't
matter since it can be proved that these weights appear in the partition
function with an even power);
the configurations (c) have weight $\sqrt 3$. This, in terms of ASM consists of
considering what is called the $3-$enumeration $A(n;3)$, i.e. an enumeration
in which 
each matrix has a weight $3^{\#(-1)}$ ($\#(-1)$ is the number of $-1$
present in the matrix). The result is
\eq
(-1)^{\frac{n(n-1)}{2}} 
S_{Y_n} (1, .., 1, -1, .., -1) = A(n;3).
\en
The explicit formula for the $3-$enumeration is \cite{kuperberg, kuperberg2}
\eq\label{3enum}
\begin{split}
A(2n+1;3) &= 3^{n(n+1)}\left(
\prod_{i=1}^n\frac{(3i-1)!}{(n+i)!} \right)^2\\ 
A(2n;3) &= 3^{n-1} \frac{(3n-1)!(n-1)!}{(2n-1)!^2}A(2n-1;3).
\end{split}
\en
\eq
\Psi_{0, n}(1, .., 1) = A(n;3) = 1, 2, 9, 90, 2025, 102060, 11573604, ...
\en

The ratio of eq.(\ref{sol-Psi1}) and eq.(\ref{sum=1}) gives the probability of
formation of the maximally nested diagram, 
which is in accord with the numerical values computed in the
first few cases.

\sk

Let us pass to the other maximally nested components. First we fix the
notation: for us a MNC of kind $m, k$ is made of two diagrams of nested
arcs. The first one has arcs connecting the pairs $(i, 1-i)$ while
the second one has arcs connecting the pairs $(m + i, m + 1 - i)$ (with
the periodic identification $i \sim i + 2(m+k)$).
\begin{center}
\includegraphics{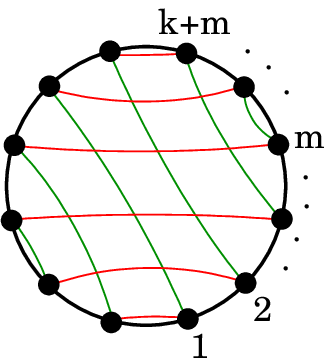}
\end{center}
For convenience we rename our spectral parameters in the following way: for
$1 \leq i \leq m$ $z_i = y_i$ and $z_{i+m+k} =\tilde y_{m+1-i} $, while for
$1 \leq i \leq k$ $z_{i+m} = x_{k+1-i}$ and $z_{i+k+2m} = \tilde x_{i}$.
We call the corresponding component $\Psi^{(0)}_{m,k}(x_1, .., x_k; \tilde x_1,
.., \tilde x_k; y_1, .., y_m; \tilde y_1,
.., \tilde y_m;)$. The role of the superscript $(0)$ will be apparent in a
moment.  
In order to determine such a component we use the exchange equation to
derive from eq.(\ref{propor3}) a recursion relation in $k$ 
(or equivalently in $m$), which is easily solved once one knows the solution
of eq.(\ref{recur-tilde}).

Let us introduce a slightly more general family of objects which we call
$\Psi^{(j)}_{m,k}$. 
They are defined for each value of the superscript $0\leq j \leq m$ and, for
a given $j$, $\Psi^{(j)}_{m,k}$ corresponds to the component  
made of  a pair of diagrams where the first has arcs connecting the
pairs $(i, 2(m+k)-i+1)$, while the second has an arc connecting $m-j$ and
$m-j+1$, arcs connecting  the pairs $(m + i, m + 1 - i)$ for $ m +j+1 < i
\leq m+2k$, and  for $0 < i  \leq  j$ arcs connecting  the pairs $(m +1- i,
m + 2 - i)$. The choice of the variables is made as in the following picture
\begin{center}
\includegraphics{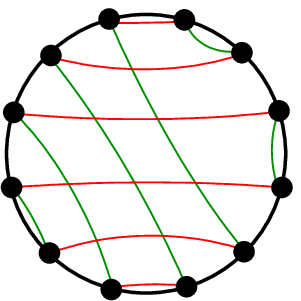}
\put(-60,-9){$\tilde x_k$}
\put(-85,8){$\tilde x_1$}
\put(-98,30){$\tilde y_1$}
\put(-58,90){$\tilde y_m$}
\put(-15,74){$y_m$}
\put(2,55){$x_k$}
\put(2,30){$y_{m-j}$}
\put(-30,-7){$y_1$}
\end{center}
The exchange equation allows us to write $\Psi^{(j)}_{m,k}(.., x_k; .. y_{m-j},
..)$ in terms of the same quantity with $x_k$ and $y_{m-j}$ exchanged and of
$\Psi^{(j\pm 1)}_{m,k}$ 
\eq
\begin{split}
q(x_k- q y_{m-j})(x_k + q y_{m-j}) \Psi^{(j)}_{m,k}(.., x_k; .. y_{m-j},
..)&\\=   q(y_{m-j} + q x_{k})(y_{m-j} - q x_{k})\Psi^{(j)}_{m,k}(.., y_{m-j}&;
.. x_k, 
..)\\ + (x_{k} - y_{m-j})(q 
x_{k} +y_{m-j}) \Psi^{(j+1)}_{m,k}(.., x_k&; .. y_{m-j},
..)\\ + (x_{k} - y_{m-j})(q 
x_{k} +y_{m-j}) \Psi^{(j-1)}_{m,k}(.., y_{m-j+1}; .. y_{m-j-1}, x_k&, y_{m-j},
..),
\end{split}
\en
this is nothing else than a particular case of eq.(\ref{exch2}).

If we now set $\tilde x_k = -q x_k $ we have that the term
$\Psi^{(j)}_{m,k}$ with $x_k$ and  
$y_{m-j}$ exchanged and the term $\Psi^{(j-1)}_{m,k}$ are zero. Then
precisely at the point $\tilde x_k = -q x_k$ we have the simplification
\eq
\Psi^{(j)}_{m,k}(.., x_k; .. y_{m-j},..) =\frac{(x_{k} - y_{m-j})(q 
x_{k} +y_{m-j})}{q(x_k- q y_{m-j})(x_k + q y_{m-j})} \Psi^{(j+1)}_{m,k}(..,
x_k; .. y_{m-j}, ..),
\en 
\begin{center}
\includegraphics{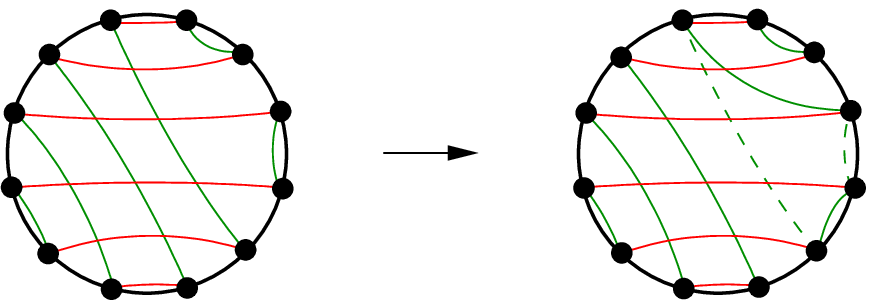}
\put(0,53){$y_{m-j}$}
\put(2,30){$x_k$}
\put(-15,5){$y_{m-j-1}$}
\put(-165,50){$x_k$}
\put(-165,30){$y_{m-j}$}
\put(-180,5){$y_{m-j-1}$}
\end{center}
hence we can write $\Psi^{(0)}_{m,k}$ in terms of $\Psi^{(m)}_{m,k}$
\eq\label{m-0}
\Psi^{(0)}_{m,k}(.., \tilde x_k = -q x_k;..) =\frac{\prod_{j=1}^m(x_{k} -
  y_{j})(q  
x_{k} +y_{j})}{\prod_{j=1}^mq(x_k- q y_{j})(x_k + q y_{j})}
\Psi^{(m)}_{m,k}(.., \tilde x_k = -q x_k; ..).
\en 
But now we notice that we can apply to $\Psi^{(m)}_{m,k}(.., \tilde x_k = -q
x_k; ..)$ the recursion relation in eq.(\ref{propor3})
\eq\label{recur-m-0}
\Psi^{(m)}_{m,k}(.., \tilde x_k = -q x_k; ..) = \prod_{z\neq x_k, \tilde
  x_k} q(x_k -q z)(x_k + 
q z) \Psi^{(0)}_{m,k-1}(.., \hat{\tilde x}_k; .., \hat{x}_k; ..) 
\en
Combining eq.(\ref{m-0}) and eq.(\ref{recur-m-0}) we get to
\eq\label{recursion3}
\Psi^{(0)}_{m,k}(.., \tilde x_k = -q x_k;..) = \prod_{j=1}^m(x_{k} -
  y_{j})(q  
x_{k} +y_{j})  \prod_{z\neq x_k, \tilde
  x_k, y_i} q(x_k -q z)(x_k + 
q z) \Psi^{(0)}_{m,k-1}(.., \hat{\tilde x}_k; .., \hat{x}_k; ..)
\en
which is the recursion relation searched. 
\begin{center}
\includegraphics{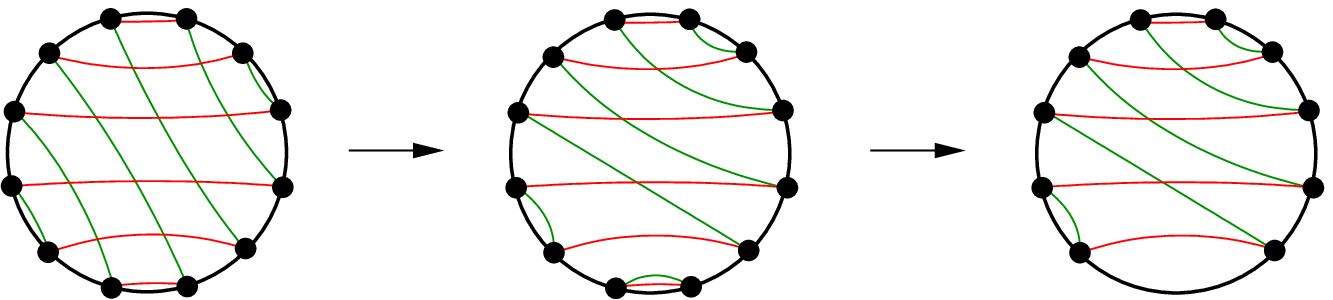}
\put(-295,55){$y_{m}$}
\put(-310,75){$x_k$}
\put(-330,-6){$y_1$}
\put(-405,-8){$-q x_k = \tilde x_k$}
\put(-163,75){$y_{m}$}
\put(-150,55){$y_{m-1}$}
\put(-163,5){$y_1$}
\put(-180,-6){$x_k$}
\put(-255,-8){$-q x_k = \tilde x_k$}
\put(-13,75){$y_{m}$}
\put(-0,55){$y_{m-1}$}
\put(-13,5){$y_1$}
\end{center}

Since in the following we will be
concerned only with $\Psi^{(j)}_{m,k}$ with $j=0$, we will no longer write the
superscript $(0)$, intending $\Psi_{m,k}   = \Psi^{(0)}_{m,k}$ . 
To proceed further we extract from $\Psi_{m,k}$ the trivial factors
introducing the function  $\tilde\Psi_{m,k}$
\eq
\begin{split}
\Psi_{m,k}(x; \tilde x; y; \tilde y) = \prod_{0<i<j\leq k}(x_j+q
x_i)(\tilde x_i + q \tilde y_j)\prod_{0<i<j\leq m}(y_i + q y_j)(\tilde y_j +
q \tilde y_i)\\\times\prod_{\substack{0<i\leq k\\0<j\leq m}}(x_i+ q \tilde
y_j)(y_j+ q x_i)(\tilde x_i + q y_j)( y_j + q \tilde
x_i)~\tilde\Psi_{m,k}(x; \tilde x; y; \tilde y).
\end{split} 
\en
$\tilde\Psi_{m,k}$ is a polynomial separately symmetric in the $x$, $\tilde
x$, $y$ and $\tilde y$. It is of degree at most $k-1$ in each $x$ or
$\tilde x$, while its degree in each $y$ or $\tilde y$ is at most $m-1$.
The recursion relation in terms of $\tilde\Psi_{m,k}$ is very simple
\eq
\tilde\Psi_{m,k}(.., \tilde x_k = -q x_k;..) = \prod_{0<i<k}q(x_k - q
x_i)(x_k + q \tilde x_i)\tilde\Psi_{m,k-1}(.., \hat x_k; .. \hat {\tilde
  x}_k;..).  
\en
It has the same form of eq.(\ref{recur-tilde}) and the $y$ and $\tilde y$
play a spectator role, hence the solution has a simple factorized form 
\eq
\tilde\Psi_{m,k}(x; \tilde x; y; \tilde y) = q^{2k(k-1)}S_{Y_k} (x_1, .., x_k,
-\tilde x_{1}, .., -\tilde x_k)q^{2m(m-1)}S_{Y_m} (y_1, .., y_m, -\tilde
y_{1}, .., -\tilde y_{m}). 
\en
Reinserting the trivial factors and evaluating at the homogeneous point we
find
\eq
\Psi_{m,k}(1; 1; 1; 1) = A(m;3)A(k;3),
\en
which again is in accord with the numerical calculation of the eigenvector
for small sizes.
We notice that among the $\Psi_{m,k}$ with $m+k=n$ is the smallest
component $\Psi_{\lfloor n/2 \rfloor, \lfloor (n+1)/2\rfloor}$
of the eigenvector for the system of size $2n$, hence we have not only proved
the conjecture of Batchelor, de Gier and Nienhuis \cite{rotor}, but we have
also determined the value of a whole family of components which contains the
smallest one.

\subsection{Other periodic boundary conditions}

We consider now systems on a cylinder (PBC) of odd or even size, where loops
wrapping 
around the cylinder are not allowed to contract.
Since these two cases are quite similar in structure we treat them together
and use an uniform notation. A maximally nested component will be denoted by
$\Psi^*_{m,k}$, where $m+k= 2n+1$ or $2n$ depending on the size of the
system. In the first case the MNC has red arches connecting points $i$ and 
$1-i$, from the point $(m+k+1)/2= n+1$ starts an unmatched red line; while the
green arches connect points $m+i$ and $m+1-i$, and there is a green
unmatched lines emanating from the point $m+n+1$ (with always the
identification $i \sim 2n+1+i$).
For the system of size $m+k=2n$, the MNC
consists of red arcs 
joining points $i$ and $1-i$, green arches joining points $m+i$ and $m+1-i$,
and the 
puncture (i.e. the hole of the cylinder) lies 
in the region delimited by the red arc from $n$ to $n+1$, the green arc from
$m+n$ to $m+n+1$, and the boundary of
the disk. 
\begin{center}
\includegraphics{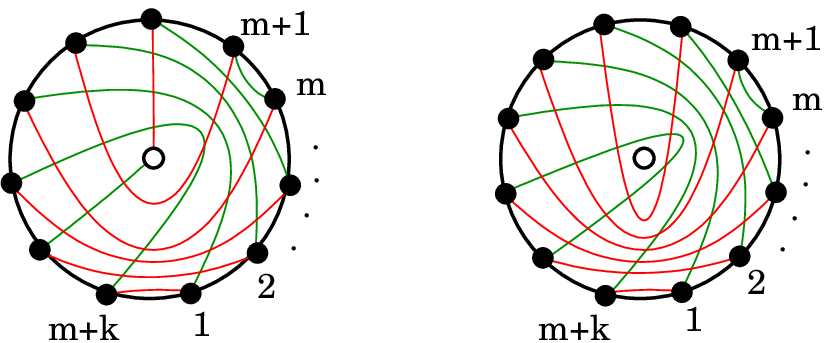}
\put(-225,-18){Odd case}
\put(-85,-18){Even case}
\end{center}

The values of these MNCs are obtained exactly
by the technique used before in the case of even size: one derives a
recursion relation for 
the components using the projection and the known
recursions of the $O(1)-$model \cite{dzz, philippe1}. Here again we stress
that the derivation of the recurrence relies on the assumption that
the degrees of the components of the rotor model are two times the
degree for the corresponding fully packed $O(1)$ model. This means that
the components of the eigenvector of the lattice of size $2n+1$ have
total degree $2n(2n+1)$, and $4n$ in each variable $z_i$. In the case of
size $2n$ the total degree is $2n(2n-1)$, and the 
degree in each variable is $4n-2$. 

Before coming to the MNCs let quickly go through the sum rule, which is simply
given by the corresponding $O(1)$ sum 
\cite{dzz, philippe1} with the variables squared 
\eq\label{somma-odd}
\somma_{2n+1}(z_1, z_2, .., z_{2n+1})  = S_{Y_n}(z_1^2,
z_2^2, .., z_{2n+1}^2) S_{Y'_n}(z_1^2,
z_2^2, .., z_{2n+1}^2). 
\en
\eq\label{somma-infty}
\somma^{(+\infty)}_{2n}(z_1, z_2, .., z_{2n})  =
S_{Y_n}(z_1^2, z_2^2, .., z_{2n}^2) S_{Y'_n}(z_1^2, 
z_2^2, .., z_{2n}^2). 
\en
Here $S_{Y_n}$ and $S_{Y'_n}$ are Schur functions. $Y_n$ is the
Young diagram having two rows of length $n-1$, two rows of length
$n-2$, .., two rows of length $2$ and two rows of length $1$. $Y'_n$
is obtained from $Y_n$ by adding one row of length $n$.  

At the homogeneous point the sums become
\eq\label{somma-odd_hom}
\somma_{2n+1}(1, 1,.., 1)  = 3^{n^2} A_{HT}(2n+1) =
3^{n^2}\prod_{j=0}^{n}\frac{4}{3} \left( \frac{(3j)!j!}{(2j)!^2}\right)^2,
\en
\eq\label{somma-infty-hom}
\somma^{(+\infty)}_{2n}(1, 1, .., 1)  = 3^{n(n-1)} A_{HT}(2n) =
3^{n(n-1)} \prod_{j=0}^{n-1}\frac{3j+2}{3j+1}\left(\frac{(3j+1)!}{(n+j)!}\right)^2.
\en
\eq\label{somma-values}
\begin{split}
\somma_{2n+1}  = 1, ~3\cdot 3, ~3^4\cdot 25, ~3^9\cdot 588, \dots\\
\somma^{(+\infty)}_{2n}  =  2, ~3^2\cdot 10, ~ 3^6 \cdot 140, \dots ;
\end{split}
\en

We return now to the MNCs and as before we begin by deriving the parallel
one, which in our notation is $\Psi^*_{0,k} $.
In order to obtain the MNC, we first extract all
the trivial factors using eq.(\ref{symmetric1}). 
\eq\label{factor-odd}
\Psi^*_{0,k}(z_1, .., z_{k}) = \prod_{1 \leq i < j
  \leq k} (z_i + q z_j)~\tilde\Psi^*_{0, k}(z_1, .., z_{2n+1});
\en

%
%

The remaining
nontrivial factors are symmetric polynomials whose degree in each
$z_i$ are $k-1$.
Then, using the recursion 
relation for the full component, we find a recursion 
relation for the nontrivial factors; these read in the two cases

%
%
%
%
\eq\label{ricorsione-odd}
\tilde\Psi^*_{0,k}(z_1 = -q^2 z_{k}, .., z_{k}) =
\frac{q^{-k}}{q-q^2} z_{k} \prod_{j=2}^{k-1} (q^2 z_{k}^2 - z_j^2)
\tilde\Psi^*_{0,k-2}(z_2, .., z_{k-1})
\en

%


Eq.(\ref{ricorsione-odd}) 
allows for a unique
solution of given degree. In the case PBC+$\infty$, i.e. $k=2n$, the
solution is
\eq\label{sol-odd}
\tilde\Psi^*_{0,2n}(z_1, .., z_{2n}) = (3q)^{-n} \frac{\prod_{i \neq
    j}(z_i + q z_j)}{\prod_{i<j}(z_i - z_j)}~ {\rm \Large Pf}~
\left[\frac{z_i^2 - 
  z_j^2}{(z_i + q z_j)(z_j + q z_i)}\right]_{i,j};
\en
while the odd case, $\tilde\Psi^*_{0, 2n+1}$ is obtained from
$\tilde\Psi^*_{0, 2n+2}$ 
\eq\label{sol-infty}
\tilde\Psi^*_{0,2n+1}(z_1, .., z_{2n+1}) = (-q)^{-n}\lim_{z_{2n+2}\to
  \infty}\frac{\tilde\Psi^*_{0, 2n+2}(z_1, .., z_{2n+2})}{(z_{2n+2})^{2n+1}}. 
\en

Unfortunately we are not able to recognise these polynomials as
partition functions of some inhomogeneous vertex model.
Nonetheless when we consider the full components at the homogeneous
point ($z_i=1$), they give
\eq\label{hom-odd}
\begin{split}
\Psi^*_{0,2n+1}(1, .., 1) &= ~1, ~ \frac{5}{3},~ \frac{127}{3^2},~
\frac{16364}{3^3}, ~\dots~~\\ \Psi^*_{0, 2n}(1, .., 1) &= ~
\frac{2}{3},~ \frac{22}{3^2},~ 
\frac{1244}{3^3},~ \frac{358312}{3^4}, ~\dots~~  
\end{split}
\en


It is not difficult to compute, in the same way as done in the previous
paragraph, all the other maximally nested components $\Psi^*_{m,k}$. 
One finds that, once all the trivial factors are eliminated, the remaining
polynomials are symmetric separately in $m$ and $k$ variables and satisfy
again a recursion relation which is easily solved and has a factorised form
\eq
\tilde\Psi^*_{m,k}(z_1, .., z_m, z_{m+1}, .., z_{m+k})=
\tilde\Psi^*_{0,m}(z_1, .., z_m)\tilde\Psi^*_{0,k}(z_{m+1}, .., z_{m+k}).
\en 
This remains true for the full components in the homogeneous limit
\eq
\Psi^*_{m,k}(1, .., 1)=
\Psi^*_{0,m}(1, .., 1)\Psi^*_{0,k}(1, .., 1).
\en

Notice that our numerical values are not integers, but this is a consequence
of our choice of normalization of the recurrence relation. Of course what
matters are ratios (or we could renormalise everything just multiplying by
appropriate powers of $3$) and the results are in accord with the numerical
computations.

\subsection{Closed boundary conditions}

In the case of closed boundary conditions, the boundary Yang-Baxter
equation allows to show that each transformation $z_1
\to 1/z_1$ and $z_N \to 1/ z_N$ preserves the eigenvector of the
double-row transfer matrix, which is called $\Phi$ in order to distinguish it
from the eigenvector of the periodic system. Since we assume the components
to be 
polynomials 
of degree  $4(\lceil N/2\rceil-1)$ in each variable, in order to maintain the
polynomiality we must have
\eq
\Phi(z_1, \dots, z_N) = z_1^{8(\lceil N/2\rceil-1)} \Phi(z_1^{-1},
\dots, z_N) 
\en
and the same for $z_N \to 1/ z_N$

Let us come now to the evaluation of the sum of the components. As in the
previous sections the mapping to the $O(1)-$model gives straightforwardly the
result \cite{philippe2, paul}.  
\eq\label{sumCBC}
\somma^{\rmCBC}_{N}(z_1, .., z_{N})  = \somma^{\rmCBC-O(1)}_{N}(z_1^2,.., 
z_{N}^2) = \chi_N(z_1, .., z_N)
\en
where $\chi_N$ is a character of the symplectic group and is defined as follows
\eq\label{chi_def}
\chi_N(z_1, .., z_N) = \left(\prod_{i=1}^N z_i^{4(\lceil N/2\rceil-1)}\right) \frac{\det(z_i^{j+ \lceil j/2
    \rceil -1} -z_i^{-j- 
    \lceil j/2\rceil +1})_{1 \leq i,j \leq N}}{\det(z_i^{j} -z_i^{-j})_{1
    \leq i,j \leq N}}  
\en
For lattices of even size \cite{philippe1}
\eq\label{sumCBCeven}
\somma^{\rmCBC}_{2n}(z_1, .., z_{2n})  = \chi_{2n}(z_1^2,.., 
z_{2n}^2) = Z^{\rmVASM}_n(z^2_1, .., z^2_{2n})
\en
where $Z^{\rmVASM}_n$ 
is the partition function of the inhomogeneous six-vertex model with
DWBC and vertical symmetry, which at the homogeneous point reduces to
the enumeration of vertically symmetric ASM \cite{philippe1, kuperberg2},
\eq\label{Uenumeration}
Z^{\rmVASM}_n(1, .., 1) = 3^{n(n-1)} A_V(n; 1) = 3^{n(n-1)} \prod_{j=0}^{n-1}
(3j+2)\frac{(2j+1)!(6j+3)!}{(4j+2)!(4j+3)!} =
\en
$$
1, ~~3^2\cdot 3, ~~3^6\cdot  26, ~~3^{12}\cdot 646, ~~3^{20}\cdot 45885,
~~3^{30}\cdot 9304650,  
$$


\noindent
For systems of odd size
we get
\cite{philippe2}
\eq 
\somma^{\rmCBC}_{2n+1}(1, .., 1)= \chi_{2n+1}(1, .., 
1) = 3^{(n-1)^2} \sN_8(2n) =  3^{(n-1)^2} \prod_{j=0}^{n-1}
(3j+1)\frac{(2j)!(6j)!}{(4j)!(4j+1)!} = 
\en
$$
1, 6, 891, 3346110, 319794090309
$$
where $\sN_8$ is the number of cyclically symmetric transpose complement
plane partitions.

\sk

We discuss now two three kinds of maximally nested components. When
the size of the system is $2n$, the MNC $\Phi_{2n}$ has arcs  of both
colours joining points $i$ and $2n+1-i$.
\begin{center}
\includegraphics{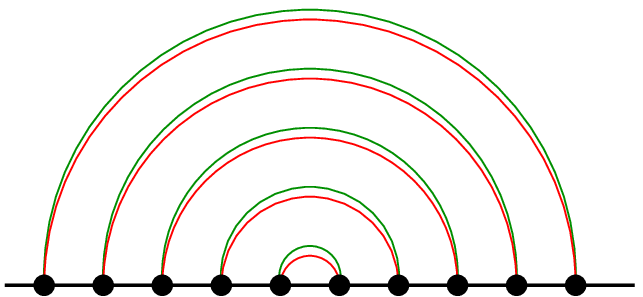}
\put(-180,-10){$z_1$}
\put(-108,-10){$z_n$}
\put(-90,-10){$z_{n+1}$}
\put(-25,-10){$z_{2n}$}
\end{center}
For odd 
lattice size $2n +1$ we consider two kinds of MNC. First $\Phi_{2n+1}^{(a)}$
which has arcs of both colours joining the pair $(i,2n+1-i)$ and two
unmatched lines emanating from the rightmost point $2n+1$.
\begin{center}
\includegraphics{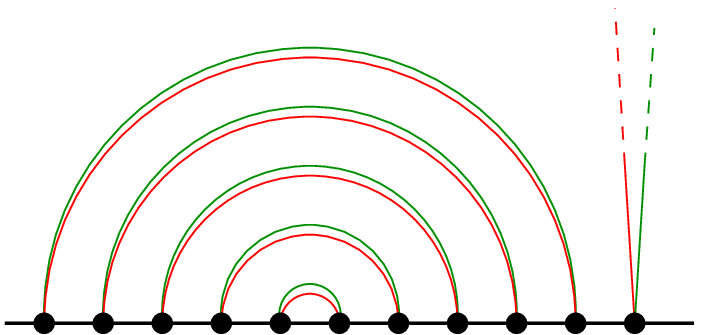}
\put(-198,-10){$z_1$}
\put(-128,-10){$z_n$}
\put(-108,-10){$z_{n+1}$}
\put(-25,-10){$z_{2n+1}$}
\end{center}
Then $\Phi_{2n+1}^{(s)}$ which has red arcs joining $(i,2n+1-i)$, green arcs
joining $(i+1,2n+2-i)$, a red line emanating from $2n+1$ and a green line
emanating from $1$.  
\begin{center}
\includegraphics{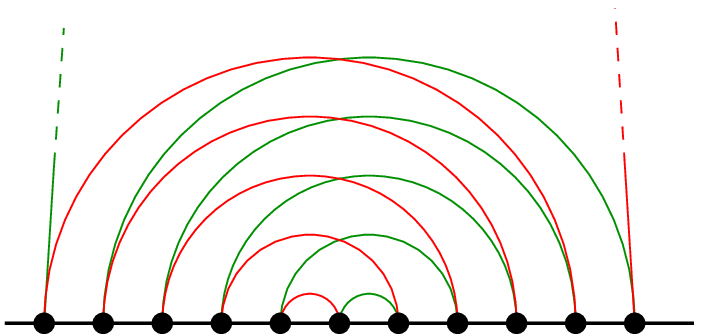}
\put(-198,-10){$z_1$}
\put(-130,-10){$z_n$}
\put(-113,-10){$z_{n+1}$}
\put(-25,-10){$z_{2n+1}$}
\put(-88,-10){$z_{n+2}$}
\end{center}
This last component is also conjectured to be the smallest one for the
ground 
state of the system with odd size \cite{rotor}.

The polynomials corresponding to this components contain a lot of trivial
factors, determined not only by the exchange relation as before, but also by the
transformation properties under  $z_1 \to 1/z_1$ and $z_N \to 1/z_N$ (with
$N=2n, 2n+1$)
\eq\label{factor-open-even}
\Phi_{2n}(z_1, .., z_{2n}) = \!\!\!\!\!\prod_{1\leq i<j \leq
  n}\!\!\!\!\! (z_i + q 
z_j)(1 + q z_i z_j) \!\!\!\!\!\!\!\prod_{n+1\leq i<j \leq 
  2n}\!\!\!\!\!\!\! (z_i + q z_j)(z_i z_j + q )~
\tilde\Phi_{2n}(z_1, .., z_{2n}) 
\en
\eq\label{factor-open-odd-a}
\Phi_{2n+1}^{(a)}(z_1, .., z_{2n+1}) =\!\!\!\!\! \prod_{1\leq i<j \leq
  n} \!\!\!\!\!(z_i + q 
z_j)(1 + q z_i z_j) \!\!\!\!\!\!\!\!\!\!\prod_{n+1\leq i<j \leq 
  2n+1}\!\!\!\!\!\!\!\!\!\! (z_i + q z_j)(z_i z_j + q
)~\tilde\Phi_{2n+1}^{(a)}(z_1, .., z_{2n+1}) 
\en
\eq\label{factor-open-odd-s}
\Phi_{2n+1}^{(s)}(z_1, .., z_{2n+1}) =\!\!\!\!\! \prod_{1\leq i<j \leq
  n+1} \!\!\!\!\!(z_i + q 
z_j)(1 + q z_i z_j) \!\!\!\!\!\!\!\!\!\!\prod_{n+1\leq i<j \leq 
  2n+1}\!\!\!\!\!\!\!\!\!\! (z_i + q z_j)(z_i z_j + q
)~\tilde\Phi_{2n+1}^{(s)}(z_1, .., \hat z_{n+1}, z_{2n+1}) 
\en
The nontrivial factors $\tilde\Phi_{2n}$ and $\tilde\Phi_{2n+1}^{(a)}$ are
polynomial of degree $2(\lceil N/2\rceil-1)$ in each variable, separately 
symmetric in the first $n$ and second $n$ (or $n+1$) variables.
The polynomial $\Phi_{2n+1}^{(s)}$ instead, has become independent of the
variable $z_{n+1}$. It is of degree $n-1$ in each variables and separately
symmetric in the first and last $n$ variables, exactly as $\tilde\Phi_{2n}$.
Moreover all these polynomials 
inherit from the respective $\Phi$ the behaviour under the change $z_{1} \to
1/z_{1}$ 
and $z_{N} \to 1/z_{N}$.

Let us analyse first the even case. It is no surprise that we can write a
recursion relation for $\Phi$ and 
hence for $\tilde\Phi$
\eq\label{rec_phi}
\tilde\Phi_{2n}(z_{n+1} = -q^2 z_n) =
q^{n-1} \prod_{i=1}^{n-1}(q z_n -z_i)(q z_n z_i -1) \prod_{i=n+2}^{2n}(q z_n
+z_i)(q z_n z_i +1) \tilde\Phi_{2n-2}(.., \hat{z}_n; \hat{z}_{n+1},
..)   
\en
The unique solution of this equation of degree $2(\lceil N/2\rceil-1)$ is  
\eq
\chi_{2n}(-z_1, .., -z_n, z_{n+1}, ..,z_{2n}).
\en
This allows us to identify the homogeneous limit of the MNC. 
One finds again a $3-$enum\-er\-ation, but this time of vertically
symmetric ASMs 
\eq
\begin{split}
\chi_{2n}(-1, .., -1, 1, ..,1) = Z^{\rmVASM}_n(-1, .., -1, 1, ..,1) =  A_V(2n + 1; 3) =\\
\frac{3^{\frac{n(n-3)}{2}} }{2^n} \prod^n_{j=1} \frac{(j-1)!(3j)!}{j
  (2j-1)!^2} = 1, 5, 126, 16038, 10320453, ..
\end{split}
\en
$A_V(2n+1; 3)$ is
the $3$-enumeration of VASM.

In the odd case, we derive the recursion relation for $\Phi_{2n+1}^{(s)}$
using first the exchange equation in order to reduce to a configuration with
two arcs of different colours joining $n$ and $n+1$, and then the usual
recursion relation. 
\begin{center}
\includegraphics{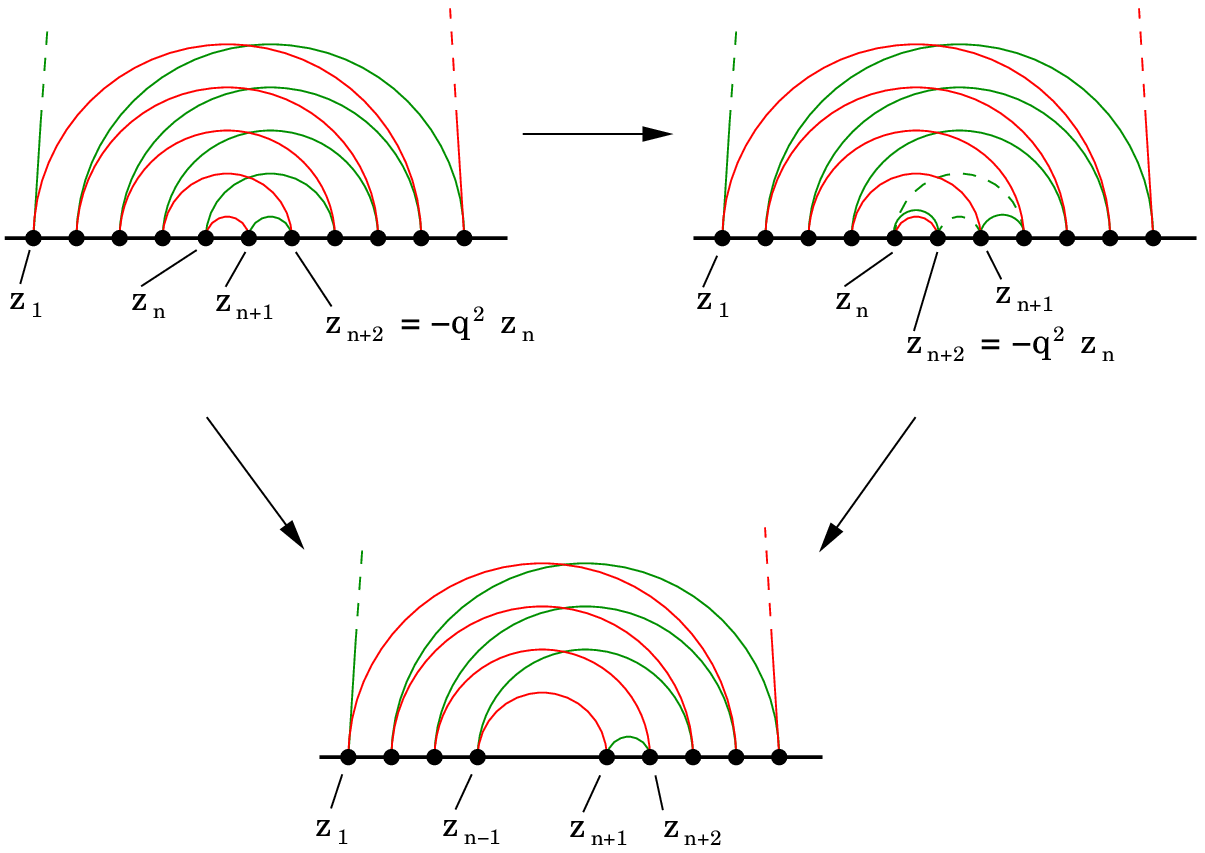}
\end{center}
Then we notice that the
recursion relation so obtained has exactly the same form as the one for
$\Phi_{2n+1}^{(s)}$, 
simply relabelling the last $n$ variables $z_i \to z_{i+1}$, i.e. 
\eq\label{rec_phi-odd}
\tilde\Phi_{2n+1}^{(s)}(z_{n+2} = -q^2 z_n) =
q^{n-1} \prod_{i=1}^{n-1}(q z_n -z_i)(q z_n z_i -1) \prod_{i=n+3}^{2n+1}(q z_n
+z_i)(q z_n z_i +1) \tilde\Phi_{2n-1}(.., \hat{z}_n; \hat{z}_{n+2}.
..)   
\en
Hence the solution is simply
\eq
\tilde\Phi_{2n+1}^{(s)}(z_1, .., z_n; z_{n+2}, .., z_{2n+1}) =
\chi_{2n}(-z_1, .., -z_n, z_{n+2}, ..,z_{2n+1}) 
\en
and the homogeneous limit again $A_V(2n+1; 3)$, as stated in the conjecture
of Batchelor et al.

It remains to treat the component $\tilde\Phi_{2n+1}^{(a)}$. 
We can again write for it a recursion relation,
completely analogous to 
eq.(\ref{rec_phi}). Its unique solution is
$$
\chi_{2n+1}(-z_1, .., -z_n, z_{n+1}, ..,z_{2n}, z_{2n+1}) 
$$
In the homogeneous limit it reduces to
\eq
\begin{split}
\chi_{2n+1}(-1, .., -1, 1, ..,1) = 
\frac{3^{\frac{n(n-1)}{2}} }{2^n(2n+1)!} \prod^n_{j=1} \frac{j!(3j+1)!}{j~
  (2j-1)!^2} 
\end{split}
\en
But now notice something unexpected, namely that
\eq
\frac{1}{2^n(2n+1)!} \prod^n_{j=1} \frac{j!(3j+1)!}{j~
  (2j-1)!^2} = \prod^{n+1}_{j=1} \frac{(3j-2)!}{
  (n+j)!} = A(n+1;1),
\en
hence we get
\eq
\Phi^{(2n+1)}_{\rmMNC}(1, .., 1) =  3^{\frac{n(n-1)}{2}}  A(n+1;1) = 1,~ 2,~
3\cdot 7,~3^3\cdot 42,~ 3^6\cdot 429,~ 3^{10}\cdot 7436
\en

\section{Conclusions}

In this paper we have considered the rotor model of Martins and Nienhuis
with different boundary conditions and spectral parameters in,
the spirit of Di Francesco and Zinn-Justin. A combined use of integrability,
polynomiality of the ground state wave function and a mapping into the
fully-packed $O(1)$ model has allowed us to write a recursion relation for
the components of the ground state in the basis of pairs of link patterns. 
The main difference with respect to the $O(1)$ case is that no component is
completely factorized in trivial terms. Nonetheless we
have been able to solve the recursion relations for what we have called the
maximally nested components. 
In the homogeneous limit (all the spectral parameters $z_i$ equal to
$1$) the sum rule for different boundary conditions gives again different
$1-$enumeration of ASMs exactly as in the $O(1)$ case. On the other hand we
see the appearance of another type of enumerations namely the
$3-$enumeration, when considering the maximally nested components. In the
case of a lattice of even
horizontal size with periodic boundary conditions the maximally nested
components are given by a product of two $3-$enumerations of
ASMs. This bilinear structure remain valid for periodic systems in which we keep
track of the hole in the cylinder, but in that case we are not able to give
a combinatorial meaning to the
factors. This we think deserves further analysis.
For the case of closed boundary conditions, we have computed the smallest
component for systems of odd size, and the parallel MNC for even size,
obtaining in both cases the $3-$enumeration of VASMs.

It would be interesting to find out some other family of components. For
that purpose 
we think one should further study the exchange relations, maybe generalizing
it to a qKZ equation, letting the parameter $q$ be generic. 
A problem we see
with generic $q$ is that the mapping to the $O(1)$ model is no longer valid.
In the paper we have in different occasions mentioned a derivation of some
results which do not rely on the mapping to the $O(1)$ model and hence are
valid also 
for the qKZ equation. This allows for example to determine the recursion
relation satisfied by the solution of the qKZ
equation, obtained deforming the exchange equation of the even periodic
system. In such case one is also able to derive the level the solution.

\section*{Acknowledgements}
It is a pleasure to thank P. Di Francesco and in particular A. Sportiello and
P. Zinn-Justin for useful discussions and comments.
This work has been supported by the ANR program ``GIMP'' ANR-05-BLAN-0029-01.


\bibliographystyle{amsplain}

\end{document}